\documentclass[a4paper]{article}

\usepackage{INTERSPEECH2020}

\title{Singer Identification Using Convolutional Acoustic Motif Embeddings}
\name{Aitor Arronte Alvarez$^{12}$,Francisco G{\'o}mez-Mart{\'i}n$^1$}

\address{
  $^1$Technical University of Madrid\\
  $^2$University of Hawaii at Manoa}
  \email{arronte@hawaii.edu, francisco.gomez@upm.es }
\begin{document}

\maketitle
\begin{abstract}
 Flamenco singing is characterized by pitch instability, microtonal ornamentations, large vibrato ranges, and a high degree of melodic variability. These musical features make the automatic identification of flamenco singers a difficult computational task. In this article we present an end-to-end pipeline for flamenco singer identification based on \textit{acoustic motif embeddings}. In the approach taken, the fundamental frequency obtained directly from the raw audio signal is approximated. This approximation reduces the high variability of the audio signal and allows for small melodic patterns to be discovered using a sequential pattern mining technique, thus creating a dictionary of motifs. Several acoustic features are then used to extract fixed length embeddings of variable length motifs by using convolutional architectures. We test the quality of the embeddings in a flamenco singer identification task, comparing our approach with previous deep learning architectures, and study the effect of motivic patterns and acoustic features in the identification task. Results indicate that motivic patterns play a crucial role in identifying flamenco singers by minimizing the size of the signal to be learned, discarding information that is not relevant in the identification task. The deep learning architecture presented outperforms denser models used in large-scale audio classification problems.
\end{abstract}


\section{Introduction}

Singer identification (SID) is a research area in Music Information Retrieval (MIR) that traditionally has required significant domain adaptation \cite{zhang2003automatic, tsai2005automatic}. Previous approaches to SID have concentrated on spectral and timbral features that are sensitive to recording conditions \cite{cai2011automatic, tsai2012automatic}, on timbral and frequency-based features \cite{fujihara2010modeling}, or on a more complex combination of acoustic and high level music descriptors \cite{kroher2014automatic}. In all previous approaches considerable hand crafting of features was needed. More recently, deep learning methods have been applied in the automatic identification of Chinese  traditional singers using Mel Frequency Cepstral Coefficient (MFCC) and  long short-term memory (LSTM) networks \cite{shen2019deep}. The approach presented in this article is different. A complete pipeline for the characterization and automatic identification of flamenco singers is presented based on \textit{acoustic motif embeddings}.

Acoustic word embeddings have proven to be a very efficient representation in automatic speech recognition (ASR) applications \cite{kamper2016deep}. In ASR, an arbitrary-length speech segment is embedded into a fixed-dimensional vector. Convolutional neural networks (CNN) \cite{bengio2014word, kamper2016deep}, and LSTM \cite{chen2015query} are normally used in this context to learn whole-word embeddings from intermediate acoustic representations such as MFCCs. We derive the concept of \textit{acoustic motif embeddings} from this literature, and apply it to represent acoustic features related to the motivic patterns that characterize each individual singer. Since in a musical singing context we are not dealing directly with pre-existing units such as phones, or words, we need to extract relevant melodic motifs directly from the audio signal, and then obtain an acoustic representation from them. Fixed-length embeddings are then learned from these representations.

Extracting motivic patterns from \textit{a cappella} flamenco music recordings requires considering various aspects of the genre. On the one hand, music recordings are normally produced in noisy environments. On the other hand, the constant pitch variability with microtonal deviations, and large range of vibrato used by the same singer, makes the extraction of musically relevant patterns difficult. 

We hypothesize that small motivic patterns contain very important information related to the personal singing style that each singer has, and that these small units, reveal small singing regularities within the larger context of high pitch variability. In order to obtain patterns from these audio signals, we develop a pitch contour simplification method based on an approximation scheme that will allow us to use sequential pattern mining methods to extract relevant motifs.

The contributions of this article are the following:
\begin{itemize}
\item We propose a new approach for singer style characterization and identification based on \textit{acoustic motif embeddings}.
\item We present a motivic pattern extraction pipeline that is able to obtain singing motifs from highly irregular pitch based on a simplification contour method.
\item A hybrid neural architecture that combines residual and recurrent networks is introduced.
\item The combined approach used in this article outperforms state-of-the-art methods and shows the importance of capturing relevant information in audio signals for singing characterization.
\end{itemize}

We make software and data publicly available \footnote{https://github.com/aitor-mir/deep-flamenco}.
\section{Motivic Patterns in Flamenco Singing}
Some of the key characteristics of flamenco music such as pitch instability, the use of intervals smaller than the half-tone, the amount of variation from phrase to phrase and from singer to singer, are derived from its improvisatory nature. Melodies are characterized by a combination of short and long notes with syllabic ornamentations (\textit{melismas}), that are placed in specific locations in a phrase \cite{mora2010characterization}. Flamenco singers learn melodies belonging to different styles and acquire singing techniques by oral transmission. Even though the key elements of a particular flamenco style are passed on from singer to singer in the oral process of knowledge transmission, each performer creates significant variations of a given theme, and also from perfomance to performance. But the personal characteristics of a singer's style such as articulation, vibrato, and other phonatory traits remain. In order to capture those characteristics that differentiates one singer from another, we develop a full pipeline for SID consisting of the following components: a contour simplification method based on an approximation scheme, a sequential pattern mining algorithm that extracts the relevant motifs for each singer in the corpus, and an acoustic motif embedding architecture that learns fixed-lenght vectors from the acoustic representations of the motifs.

\subsection{Flamenco corpus}

Flamenco music, as an oral musical  culture, lacks musical transcriptions of the repertoire. The \textit{corpus}COFLA \cite{kroher2016corpus} consists of more than 1,800 music recordings taken from flamenco anthologies curated by experts following the MIR corpora research principles outlined by Serra \cite{serra2014creating}. For this article we select a sample of the \textit{a cappella} collection of \textit{corpus}COFLA  based on the minimum number of recordings available for each singer. We set this number to 3. We also make the selection of the recordings based on the variety of styles available. We concentrate on the following styles and substyles of flamenco: \textit{deblas, martinetes} and \textit{saetas} from the \textit{tonas} style, and \textit{fandangos}.

The selection brings substantial challenges related to the main goal of this research.  Stylistically, the fact that three closely related substyles of \textit{tonas} are used will make the classification task more complex. At the same time this constraint will allow us to concentrate on singer's individual differences, minimizing the effect of the style. The \textit{acoustic motif embeddings} will need to characterize the singer's particular singing style by capturing acoustic features that are unique to their voice. The descriptive statistics from Table \ref{table:motifs} gives us an idea of the regularity that some singers exhibit in their motivic improvisations (AM, EC), while others show more variability and therefore less patterns (TP, EM) .

\subsection{Contour approximation and pattern discovery}\label{contour}

From the sample of songs selected from \textit{corpus}COFLA, we develop a pipeline to extract statistically and musically relevant motifs from flamenco audio recordings characterized by high instability of pitch. The first component of that pipeline is a contour approximation method.

The first step in the pipeline is to extract the fundamental frequency $f_0$ for all raw audio signals in the collection of songs $\boldsymbol{F}$, by using a sinusoid extraction and salience function \cite{salamon2011sinusoid} with a sampling rate of 44.1 KHz and a window step of 256 samples. In previous research \cite{kroher2018audio}, contour simplifications approaches have been used to obtain robust melodic representations by converting high fluctuating frequencies into equal steps, concretely into semitones. The approach taken in this article is not determined by any pre-existing musical or tuning system. Instead, in order to be able to capture meaningful musical patterns in a microtonal improvisation, an unsupervised approach based on univariate k-means clustering \cite{qiu2007comparative} is used to group frequencies. We transform to cents all frequencies in $f_0$.

The steps  to obtain a contour approximation of all $C_i(f_0) \in \boldsymbol{C}$ for $i =  1, 2, ..., t$  where $t:=|\boldsymbol{C}|$ are:
\begin{itemize}
\item 	Given a set of points $P$ in $f_0$ we say that a line segment $L$ is bounded by all points in $P$ if $L \subseteq K_j$ and $K_j \in \boldsymbol{K}$ for $j=1, 2, ..., l$  where $l:=|\boldsymbol{K}|$ and $\boldsymbol{K}$ is the set of all k-means  clusters in $\boldsymbol{F}$. 

\item  We obtain  $C_i(f_0)=d_1, ..., d_p$ where $p$ is the number of contour elements expressed as the distance between two line segments $d(L_{j-1}, L_{j})$ and $L_{j-1} \subseteq K_{j-1}$ and $L_{j} \subseteq K_{j}$
\end{itemize}

The result is a vector of contour points represented in the time domain. We use + and – signs to specify whether the direction of the contour ascends (+) or descends (-).

Once the contour is created, we apply the BIDE algorithm \cite{wang2004bide} to extract closed sequences of contours. These sequential patterns are what we call motifs. We obtain $|\boldsymbol{S}|$ dictionaries of $|S|_n$ motifs $M(S_i)=\{m_{11}, m_{21}, ..., m_{ni} \}$ where, $\boldsymbol{S}$ is the set of all singers in the corpus, and $n$ the total number of motifs for the i-th singer $S$.

\begin{figure}
\includegraphics[width=\linewidth]{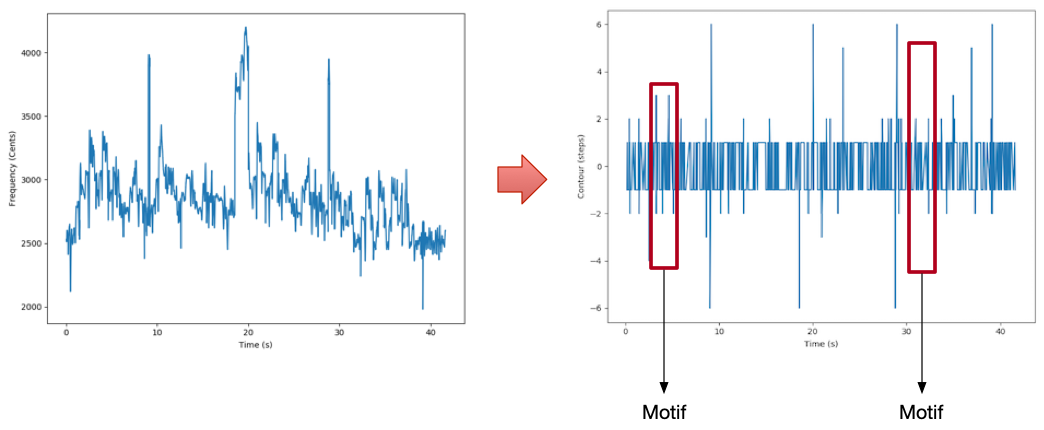}
  \caption{Contour simplification of the audio signal, from $f_0$ (left) to its contour reduction (right). Motifs are mined directly from $C(f_0)$. }
  \label{fig:contour}
\end{figure}

\section{Acoustic Embedddings}
From the dictionary $M$ of motifs, acoustic features are obtained that later will be used as the input of the different convolutional architectures. The melodic contour is used to extract relevant melodic features that may pertain to the singer's musical style, and the acoustic features to the singer's phonic characteristics. It can be argued that the acoustic feature could be associated with the quality, environmental acoustics and conditions of the recording instead of the singer's individual phonation. The fact that several different recordings are used for each singer will minimize this effect, also known as the album effect.

Following notation from \cite{kamper2016deep} we formally define a vector of frame-level features (acoustic features) as $Y= \boldsymbol{y}_{1:T}$ where each $\boldsymbol{y}_t \in\mathbb{R}^d$ is a $d$-dimensional feature at the frame level. An acoustic embedding is then a function $f(Y)$ that maps a variable length segment into a fixed-dimensional space $\mathbb{R}^d$. We say that the embeddings $f(y_1) \approx f(y_2)$ if $||f(y_1)-f(y_2) ||\geq \theta$, where $\theta$ is a minimum acceptable similarity threshold of the embeddings.

\subsection{Acoustic features}

MFCC features have been widely used in speaker recognition systems in particular, and in ASR tasks in general. SID approaches to acoustic modeling have also followed this path in polyphonic music \cite{mesaros2007singer, lagrange2012robust} and in \textit{a cappella} singing in combination with other general descriptors related to style, such as vibrato and melodic contour \cite{kroher2014automatic}. Acoustic features in MIR tasks highly depend on the quality of the recordings, and normally, other auxiliary general descriptors are needed. In the approach presented no general features or descriptors are used, and embeddings are learned directly from the acoustic representations.

Recent approaches to ASR based on deep learning methods have explored the use of spectrograms in the acoustic modeling of speech signals, improving the results of log Mel-scale filter banks energies (FBANK) \cite{hoshen2015speech}. Although such representations are used, MFCC, and in particular FBANK , tend to perform better than spectrograms in language and dialect identification tasks \cite{shon2018convolutional}. These results seem to indicate that spectrograms features work better in deep learning architectures when the time domain structure of the signal is particularly sensitive to the goal of the task. When analyzing music audio signals using deep learning networks, mel-spectrograms have proven to be a more efficient acoustic representation precisely because of the time-dependent nature of music signals \cite{choi2016automatic, choi2017convolutional}.

In the case of flamenco singning, where environmental conditions are noisy and the phonation properties of the singer are important, it is unclear what type of acoustic feature would work best with deep networks. For that reason, we test our architectures on spectrogram, log mel-spectrograms and MFCC features.
\subsection{Convolutional architectures}\label{convarch}

  We frame the learning of \textit{acoustic motif embeddings} as an acoustic representation problem. Based on the acoustic features obtained from the melodic contours, the architectures proposed in this subsection will learn the acoustic representation that will be used to characterize individual singers. The quality of the embeddings obtained will be tested on a singer classification task.
  
As a baseline, we use an architecture developed for automatic music tagging based on Convolutional Recurrent Neural Networks (CRNN) \cite{choi2016automatic}. Other CNN architectures, initially developed for image classification problems, have shown transferability to the audio domain \cite{hershey2017cnn}. For the present research, we test whether that transfer also applies to music, more specifically to singing, and use a deep residual network as the second baseline model applied in the beforementioned study on large scale audio classification. We present as a contribution, a hybrid model that combines residual layers \cite{he2016deep} with Bidirectional Long Short-Term Memory (BLSTM) networks \cite{graves2005framewise} that can capture longer sequential relations in the data, and that is also capable of better mapping the acoustic feature input to a lower dimensional space.

The CRNN architecture is composed of 4 blocks that contain a convolutional layer, a batch normalization step, an exponential linear unit layer, a maxpool layer, and a dropout layer. The first block of layers uses convolutional filters of sizes 3x3 with stride 2x2, also used in the maxpool layer. The rest of the blocks (2-4) use small filters of 3x3 with stride of 1 to capture small regularities in the data. The architecture uses a recurrent GRU network to capture the sequential nature of music signals. We obtain the vector embeddings from this last layer.

Following the recent success of deep residual networks in audio classification tasks \cite{hershey2017cnn}, we adapt the deep residual network \textit{Resnet50} to this particular domain. We make modifications to the original configuration by removing the stride of 2x2 in the first convolutional layer, and reducing the size of the first convolutional filter from 7x7 to 3x3.

The hybrid model uses shallow residual blocks present in Resnets as a front-end to process acoustic features, and a recurrent BLSTM network to learn sequential characteristics present in music audio data. We parameterize the first convolutional layer of the first residual block with 3x3 filters and stride of 2x2, and for the remaining convolutional layers in all blocks a filter with kernel size of 3x3 and stride of 1. The recurrent layer learns from an input sequence $X=\{x_1, x_2, ..., x_T\}$ the best representation that produces as output the sequence $Y=\{y_1, y_2, ..., y_T\}$, where $X$ is a vector of acoustic features at the frame level. The BLSTM is composed of a forward LSTM $\overrightarrow{f}$ that estimates the forward hidden states $\overrightarrow{h_1}, \cdots , \overrightarrow{h_T}$. The backward LSTM $\overleftarrow{f}$ obtains a backward representation of the hidden states by processing the sequence in reverse order, obtaining the backward hidden states iterating back from $t = T$ to 1. The concatenation of the output of the forward and backward networks $\overrightarrow{Y} \oplus \overleftarrow{Y}$ produces the embedding of a given motif. Figure \ref{convarch} shows the characteristics of the proposed architecture.

All architectures use a fully connected layer with 1024 units for the classification of the data instances. All models were implemented using the library Pytorch \cite{paszke2019pytorch}.

\begin{figure}
\includegraphics[width=\linewidth]{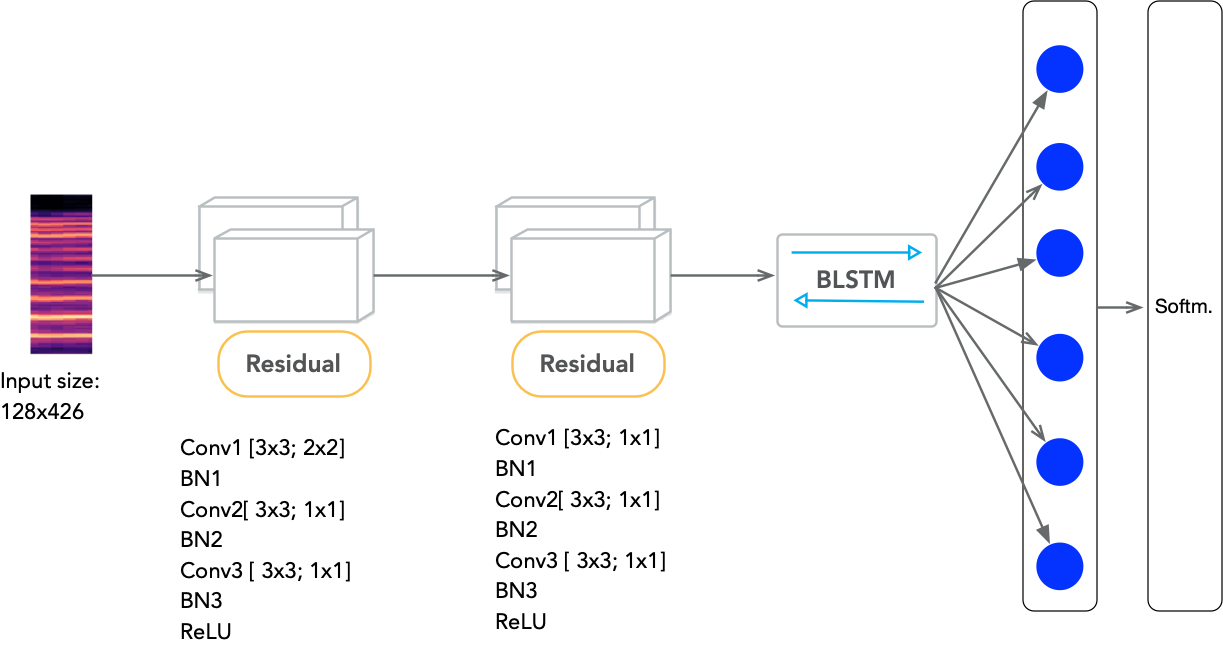}
  \caption{Hybrid Residual and BLSTM architecture }
  \label{fig:contour}
\end{figure}
\begin{table}
\centering
\caption{Motivic patterns by singer}
\label{table:motifs}
\begin{tabular}{ |c||c | |c |}
 \hline
 \multicolumn{3}{|c|}{\textbf{Motifs} } \\
 \hline
\textbf{ Singer}   & \textbf{Instances} & \textbf{Label}\\
 \hline
 Enrique Morente   & 1,125 &EM \\
 Antonio Mairena &   3,467 &AM \\
 El Chocolate & 3,064 &EC\\
Juan Talega    &2,445 &JT \\
Tomas Pavon&   1,062 &TP \\
 \hline
\end{tabular}

\end{table}

\section{Experimental Methodology}

In order to test the quality of the embeddings learned by the architectures described in \ref{convarch}, and to test more generally how \textit{acoustic motif embeddings} are able to characterize individual singer's singing styles, we perform a classification task based on identifying the singer given the motivic corpora generated in \ref{contour} for each of the two acoustic features and combinations of them. We also test how acoustic embeddings generated following the steps described in \ref{contour} improve overall identification accuracy when compared to acoustic embeddings that were randomly sampled from the raw audio signal, and with audio samples that contain the motif in a larger sample (total of 3 seconds).
\begin{table*}[h]
\label{results}
  \centering
  \begin{tabular}{ c c |c c | c c| c c }
 \hline
  &  & \textbf{MFCC}  & &   \textbf{Spectrogram}   & &\textbf{Mel-spec.} &  \\
 \textbf{Model}  & \textbf{Dataset} &\textbf{Accuracy (\%)}& \textbf{AUC} & \textbf{Accuracy (\%)} & \textbf{AUC}& \textbf{Accuracy (\%)} & \textbf{AUC}  \\
 \hline
 CRNN  & Motifs & 88.59 & 87.66 & 87.13& 86.54 & 91.05 & 89.11 \\
 Residual50 &Motifs &  87.31 & 86.55 & 83.47& 81.98 & 93.19 & 91.77 \\
 Residual-BLSTM &Motifs  & \textbf{94.78} & \textbf{92.02} & \textbf{91.62}& \textbf{88.66} & \textbf{96.29} & \textbf{94.42}  \\
 \hline
  CRNN  & Motifs+segment & 86.48 & 85.12 & 84.26 & 83.29 & 90.33 & 89.03 \\
 Residual50 &Motifs+segment &  85.01 & 84.32 & 82.41& 81.11 & 91.89 & 90.25 \\
 Residual-BLSTM &Motifs+segment  & \textbf{91.88} & \textbf{90.02} & \textbf{89.98}& \textbf{88.02} & \textbf{95.23} & \textbf{93.36}  \\
 \hline
  CRNN  & Random & 81.11 & 80.21 & 78.33 & 76.95 & 82.82 & 81.76 \\
 Residual50 &Random &  81.31 & 80.25 & 79.58 & 78.92 & 84.89 & 82.52 \\
 Residual-BLSTM &Random & \textbf{82.19} & \textbf{84.26} & \textbf{81.73}& \textbf{80.32} & \textbf{85.96} & \textbf{84.01}  \\
 
 \hline
\end{tabular}
  \caption{Results on the SID task with the different  acoustic features and datasets}
  \label{results}
\end{table*}

\subsection{Data}

From the procedure described in subsection \ref{contour} we obtain a total dataset of 11,163 motivic patterns. Table \ref{table:motifs} describes the motivic distribution by singer. For MFCC features, we set a window of 25 ms with 40 mel coefficients. Log mel-spectrogram features were obtained using a sample rate of 16 KHz, windows of 25 ms, and 128 mel filterbanks. Spectrograms features were generated based on a window of size 25 ms at a 16 KHz, with 400 bins. We zero-pad the acoustic representations to obtain equal-size vectors (128x426) for the input layer of the architectures.

\subsection{Training}

During the pre-training stage we identify optimal batch sizes and epochs based on accuracy and cross-entropy loss. We found that for the size of the sample, batches of 64, 20 epochs with early stopping policy of 2 epochs, and a learning rate of 0.001 obtain the best results. We use for all architectures the ADAM optimizer \cite{kingma2014adam} and use as the loss function the cross-entropy loss:

\begin{equation}
loss(y, \hat{y})= - \sum y \log \hat{y}
\end{equation}
where $y$ is the probability of the true class, and $\hat{y}$ is the probability of the predicted label.

From the number of motivic instances generated from the corpus, we use a subset of 80\% of the dataset for training and the remaining 20\% for testing.

\section{Results and Discussion}

Results in Table \ref{results} highlight the 3 main findings of this research: 1) acoustic motif embeddings provide relevant information that model singer-specific styles in flamenco music. 2) Acoustic features are sensitive to the overall identification task, and seem to affect more non-recurrent architectures. 3) The proposed model outperforms previous architectures for MIR and audio classification tasks. 

Motivic patterns alone (motifs and motifs+segment datasets), result in the highest accuracy gain when compared with the dataset that contains no specific motivic patterns (Random). The mean relative accuracy improvement on the top model, when we compare datasets that contain patterns with the one that does not, is 13.5\% for MFCC features, 11\% for spectrogram, and 11.5\% for mel-spectrogram. Motivic patterns seem to encapsulate and provide important information that pertain to the melodic characteristics of the singer or \textit{cantaor}. Motivic patterns also required less training when compared with non-motivic datasets. For the most accurate model (Residual-BLTSM) only 10 epochs were required, compared to the same model with the Random set that needed 27. An strict early stopping policy of 2 helped reduced the difference in training and test to only 4\% for the best model. AUC scores show overall, the discriminative power over classification classes (singers) the proposed approach has.

The proposed architecture (Residual-BLTSM) outperforms in all features tested the CRNN baseline and the deeper residual network used in large scale audio classification. We noted that non-recurrent architectures (Residual50) are particularly sensitive to acoustic features, while recurrent ones are able to learn sequential data more accurately independently of the acoustic feature used. 

CNN are very efficient models for capturing local-level features of the audio signal. When using small kernel sizes (3x3) and strides (1x1), CNN models learn high-quality acoustic representations. In the case of flamenco music, acoustic features alone tend to carry an important amount of information about a singing style, and motivic patterns highlight higher-level melodic features that increase accuracy in the identification of a singer.

The learned representations are able to characterize specific singing styles in flamenco music, and are able to identify singers with a high degree of accuracy with the minimum amount of information. Unlike previous approaches to flamenco singer identification \cite{kroher2014automatic} where significant domain knowledge was required, our approach shows that acoustic features in combination with motivic patterns are sufficient to automatically identify singers with small audio samples. This approach has significant advantages when dealing with sparse data and incomplete signals, with numerous applications to music and multimedia retrieval with minimal information.
\section{Conclusion}

A deep learning approach to model complex and individualized singing styles from flamenco music was presented. Acoustic motif embeddings are able to capture small regularities present in flamenco singing that are sufficient to accurately identify singers in closely related  flamenco styles. The combined approach taken in this article outperforms other deep learning architectures used in large-scale audio classification and MIR tasks. The architecture presented is robust to data sparsity and noise, and is able to reduce the complexity of a signal by approximating it while preserving strong predictive power.
\section{Acknowledgments}
The technical support and advanced computing resources from University of Hawaii Information Technology Services-Cyberinfrastructure are gratefully acknowledged.

\bibliographystyle{IEEEtran}

\bibliography{mybib}

\end{document}